\documentclass{ws-procs975x65}
\newcommand{\etal}{{\it et al.}}
\newcommand{\mnras}{MNRAS}

\newcommand{\apj}{ApJ}

\newcommand{\prd}{Phys.~Rev.~D}

\begin{document}

\title{\uppercase{New tests of local Lorentz invariance and local
    position invariance of gravity with pulsars}}
\author{\uppercase{Lijing Shao}$^{1,2,*}$, \uppercase{Norbert Wex}$^{1}$,
  \uppercase{Michael Kramer}$^{1,3}$}
\address{$^1$Max-Planck-Institut f\"{u}r
  Radioastronomie, Auf dem H\"{u}gel 69, D-53121 Bonn, Germany \\
  $^2$School of Physics,
  Peking University, Beijing 100871, China \\
  $^3$Jodrell Bank Centre for Astrophysics, School of Physics
  and Astronomy, \\The University of Manchester, M13 9PL, UK \\
  $^*$E-mail: lshao@pku.edu.cn}

\begin{abstract}
New tests are proposed to constrain possible deviations from local
Lorentz invariance and local position invariance in the gravity
sector. By using precise timing results of two binary pulsars, i.e.,
PSRs J1012+5307 and J1738+0333, we are able to constrain
(strong-field) parametrized post-Newtonian parameters
$\hat{\alpha}_1$, $\hat{\alpha}_2$, $\hat{\xi}$ to high precision,
among which, $|\hat{\xi}| < 3.1\times10^{-4}$ (95\% C.L.) is reported
here for the first time.
\end{abstract}


\bodymatter

\section{Introduction}\label{sec:intro}

The Einstein equivalence principle (EEP) is a far-reaching concept in the
heart of many gravity theories. If EEP is valid, then gravitation must be a
curved-spacetime phenomenon. As a direct consequence, gravity
theories which fully embody EEP are the so-called ``metric theories of
gravity''\cite{wil93,wil06}. The validity of EEP involves three
different aspects, namely the weak equivalence principle, the local Lorentz
invariance (LLI), and the local position invariance (LPI)\cite{wil06}.

In the parametrized post-Newtonian (PPN) framework\cite{wn72,wil93,wil06}, we 
present the orbital dynamics of binary systems from a generic semi-conservative 
Lagrangian, based on which, new tests of strong-field LLI-violating PPN
parameters, $\hat{\alpha}_1$ and $\hat{\alpha}_2$, are proposed\cite{sw12}. New 
limits are obtained from small-eccentricity relativistic neutron star
(NS) white dwarf (WD) systems, PSRs J1012+5307\cite{lcw+01,lwj+09} and
J1738+0333\cite{akk+12,fwe+12}. Here we briefly summarize the analysis and
results of Ref.~\citen{sw12}. We also propose a new test of the
strong-field LPI-violating PPN parameter, $\hat{\xi}$, and get a new limit
$|\hat{\xi}| < 3.1 \times 10^{-4}$. All limits in this proceeding
contribution correspond to 95\% confidence level, and the PPN
parameters with ``hat'' represent the strong-field generalization of the 
weak-field PPN parameters (without hat). The preferred frame is assumed to be 
defined by the isotropic CMB background.

\section{Local Lorentz invariance}\label{sec:lli}

LLI violation in the gravity sector is described by $\alpha_1$ and
$\alpha_2$ in the PPN framework\cite{wn72}, and these two parameters
are constrained by various observations of geophysics, Solar
System, and pulsar timing
experiments\cite{nw72,nor87,de92,bcd96,wex00,wk07,mwt08}.  Recently in
Ref.~\citen{sw12}, we find that the effects of $\hat{\alpha}_1$ and 
$\hat{\alpha}_2$ on the binary orbital dynamics decouple and manifest 
characteristic signatures when the orbital eccentricity is small (see Fig.~1 in
Ref.~\citen{swk12} for illustrations), hence they can be constrained
individually.

Damour and Esposito-Far\`{e}se are the first to work out the effects
of $\hat{\alpha}_1$ on orbital dynamics of pulsar binaries\cite{de92}. 
After dropping $\hat{\alpha}_2$ related terms, they found
that in the limit of a small eccentricity, $\hat{\alpha}_1$ induces a 
polarization
of the orbit. The effect linearly depends on $\hat{\alpha}_1$ and the binary
velocity with respect to the preferred frame, ${\bf w}$. Due to unknown
angles, previous methods can only get probabilistic limits on $\hat{\alpha}_1$.
Ref.~\citen{sw12} demonstrates that, given a sufficiently long observing time 
span, the large periastron advance would
be able to overcome probabilistic assumptions. By utilizing the limits
of eccentricity variations, we get a robust and conservative
constraint,
\begin{equation}\label{eq:a1}
  \hat{\alpha}_1 = -0.4^{+3.7}_{-3.1} \times 10^{-5} \,,
\end{equation}
from PSR J1738+0333. It surpasses the current best limit
from LLR\cite{mwt08} by a factor of five.

In the limit of a small eccentricity, $\hat{\alpha}_2$ induces a precession of
the orbital angular momentum around ${\bf w}$. It changes the orientation
of the orbital plane with respect to the Earth.
After subtracting other potential astrophysical and gravitational
contributions, we get a combined limit from PSRs J1012+5307 and
J1738+0333\cite{sw12},
\begin{equation}
  \label{eq:a2}
  |\hat{\alpha}_2| < 1.8 \times 10^{-4} \,.
\end{equation}
This limit is still three orders of magnitude less constraining than the limit
given in Ref.~\citen{nor87}, however, it is obtained for a strongly 
self-gravitating body.

\section{Local position invariance}\label{sec:lpi}

\begin{figure}[t]
\begin{center}
\psfig{file=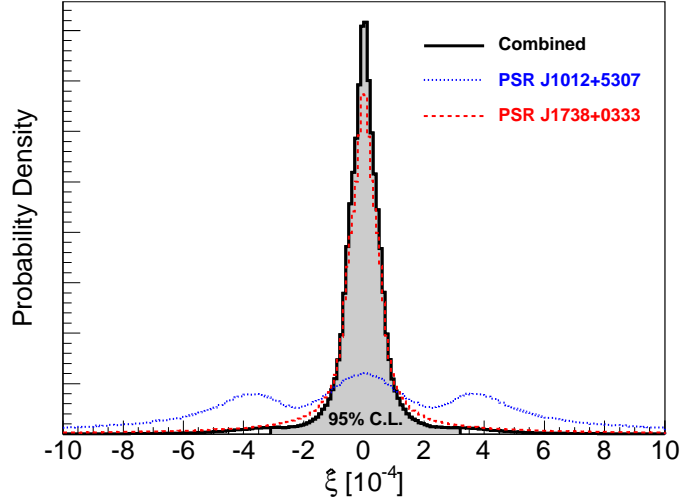,width=10cm}
\end{center}
\caption{Probability distributions of $\hat{\xi}$ from binary pulsars
  PSRs J1012+5307 (dotted blue), J1738+0333 (dashed red) and their
  combination (solid black).}
\label{fig:xi}
\end{figure}

LPI violation is described by the Whitehead's term, characterized by
$\xi$\cite{wil73,wil93,wil06}. Even for fully conservative theories of gravity 
one may have a $\xi \ne 0$. From its Lagrangian, we can immediately
identify its analogy with the $\alpha_2$ term by replacing ${\bf w}$ into
${\bf v}_G \equiv |\Phi_G|^{1/2} {\bf n}_G$ and $\alpha_2$ into
$-2\xi$, where $\Phi_G$ is the Galactic potential at the position of
the binary, and ${\bf n}_G$ is the direction of the Galactic
acceleration. Hence, for small-eccentricity binaries, $\xi$ induces a
precession of orbital angular momentum around ${\bf n}_G$, which
causes a change in the binary orientation.  The same analysis done for
$\hat{\alpha}_2$ in Ref.~\citen{sw12} applies to the $\hat{\xi}$ test. The
probability distributions of $\hat{\xi}$ from PSRs J1012+5307, J1738+0333,
and their combination are illustrated in Fig.~\ref{fig:xi} (cf. Fig.~4
in Ref.~\citen{sw12}).
From their combination, we get
\begin{equation}\label{eq:xi}
  |\hat{\xi}| < 3.1 \times 10^{-4} \,,
\end{equation}
which surpasses the limit from the non-detection of anomalous Earth
tide in gravimeter data\cite{wg76,wil06} by one order of magnitude.

\section*{Acknowledgements}

Lijing Shao is supported by China Scholarship Council (CSC).


\end{document}